# U-Net Using Stacked Dilated Convolutions for Medical Image Segmentation

Shuhang Wang*, Szu-Yeu Hu, Eugene Cheah, Xiaohong Wang, Jingchao Wang, Lei Chen, Masoud Baikpour, Arinc Ozturk, Qian Li, Shinn-Huey Chou, Constance D. Lehman, Viksit Kumar*, Anthony Samir*

*{swang38, vkumar14, asamir}@mgh.harvard.edu
Department of Radiology, Massachusetts General Hospital, Harvard Medical School, Boston, MA, USA

**Abstract.** This paper proposes a novel U-Net variant using stacked dilated convolutions for medical image segmentation (SDU-Net). SDU-Net adopts the architecture of vanilla U-Net with modifications in the encoder and decoder operations (an operation indicates all the processing for feature maps of the same resolution). Unlike vanilla U-Net which incorporates two standard convolutions in each encoder/decoder operation, SDU-Net uses one standard convolution followed by multiple dilated convolutions and concatenates all dilated convolution outputs as input to the next operation. Experiments showed that SDU-Net outperformed vanilla U-Net, attention U-Net (AttU-Net), and recurrent residual U-Net (R2U-Net) in all four tested segmentation tasks while using parameters around 40% of vanilla U-Net's, 17% of AttU-Net's, and 15% of R2U-Net's.

## 1 Introduction

Semantic segmentation is highly sought after in medical image analysis because of its capability to automate and facilitate delineation of regions of interest [1]. With the success of deep learning in recent years, many deep learning-based segmentation algorithms have been developed and applied in various medical imaging applications such as disease diagnosis [2], tissue volume estimation [3], and surgical guidance [4].

The vanilla U-Net has been widely used in medical imaging applications for both research and commercial purposes since its development in 2015 [5]. Many U-Net variants have been proposed, such as attention U-Net (AttU-Net) [6], recurrent residual U-Net (R2U-Net) [7], nested U-Net (U-Net++) [8], and 3D U-Net [9]. These methods have been shown to be effective for specific use cases, typically at greater computational resource cost. However, recent studies pointed out complicated U-Net variants might not outperform vanilla U-Net. For example, Isensee et al. [10] demonstrated a slightly modified vanilla U-Net (no new U-Net or nnU-Net) achieved competitive results on the BraTS2018 challenge for brain tumor segmentation. In another paper [11] the authors evaluated nnU-Net in the context of the Medical Segmentation Decathlon challenge, measuring segmentation performance in ten disciplines comprising distinct entities, image modalities, image geometries and dataset sizes. They reported nnU-Net achieved the highest mean Dice scores across all classes.

It has been well demonstrated the U-Net architecture is suitable for image segmentation, partly because skip connections between encoder and decoder are essential to upsample lower-resolution layers. However, the U-Net architecture introduces tradeoffs: (1) U-Net convolutions have a very limited receptive field and encoder downsampling may degrade correlation between the pixels. To obtain larger receptive fields, Devalla et al. [12] introduced dilated convolutions into U- Net, where the dilation rate was increased while the resolution was downsampled. However, Hamaguchi et al. [13] pointed out that aggressively increasing dilation rates might fail to aggregate local features due to sparsity of the kernel and potentially be detrimental to small objects. (2) Existing methods are based on an unstated assumption that the each encoder/decoder operation perceives and outputs a single receptive field at pixel, which may limit the variety of receptive fields. (3) Some medical imaging modalities, especially ultrasound, require high temporal resolution, which in turn mandates highly efficient use of computational resources, limiting the deployment of computationally intensive network architectures.

To fully take advantage of superior U-Net segmentation performance while mitigating drawbacks such as simple and small receptive fields, we propose a novel U-Net variant (SDU-Net) which processes feature maps of each resolution using multiple dilated convolutions successively and concatenates all the convolution outputs as input to the next resolution. The cascade processing increases the receptive field gradually while the concatenation gathers multiple receptive fields within the same resolution. Since the U-Net architecture processes an image in multiple resolutions, SDU-Net has the ability to sense both small and large receptive fields for either high or low resolutions. Our experiments demonstrate that SDU-Net outperforms state-of-the-art algorithms while using far fewer parameters.

## 2    Network Architecture

The proposed method, SDU-Net, adopts the overall network architecture of vanilla U-Net (Figure 1), with modification of the encoder and decoder operations (an operation indicates all the processing for feature maps of the same resolution). For each encoder/decoder operation, U-Net uses two standard convolutions and takes the output of the second convolution as the input of the next operation, while SDU-Net adopts one standard convolution followed by multiple dilated convolutions and concatenates all dilated convolution outputs as input to the next operation (Figure 2). In Figure 2, we can see the difference between encoder and decoder is that each encoder operation downsamples the preceding layer while each decoder upsamples the preceding layer and concatenates it with its corresponding encoder map (indicated by dashed-line box).

In each encoder/decoder operation, the channel number is reduced while the dilation rate is increased. To make an efficient design, we set the output channel number of each dilated convolution to $n_{out}/2$, $n_{out}/4$, $n_{out}/8$, $n_{out}/16$, and $n_{out}/16$, and therefore the concatenation output consists of $n_{out}$ channels.

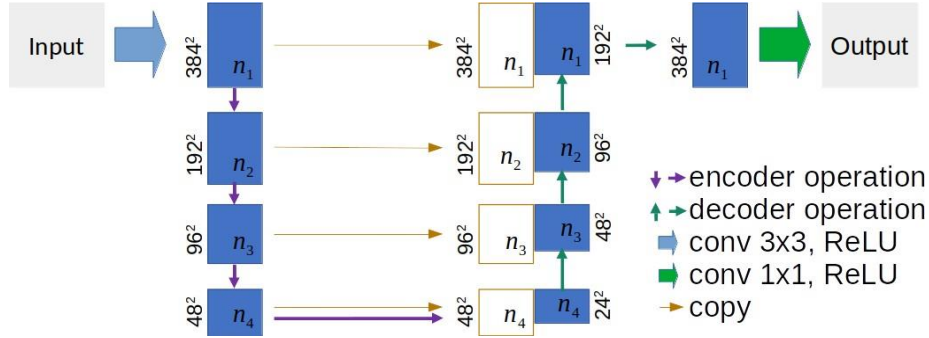

**Fig. 1.** The U-Net architecture. The boxes represent the feature maps, with the channel number denoted by equations of $n_i$ in the boxes. White boxes (brown borders) represent copied feature maps. "copy" is used to represent the copying of encoder feature map to concatenate with the decoder map in the subsequent decoder operation. The resolution is provided at the left or right edge of the box.

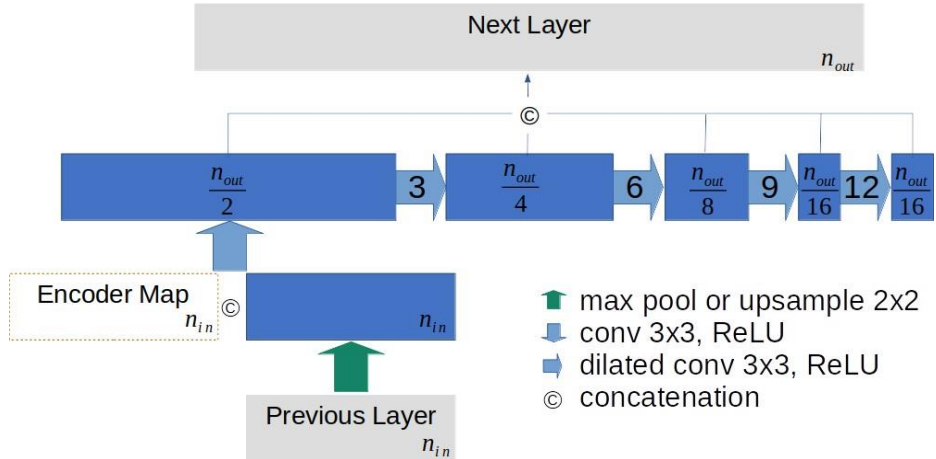

**Fig. 2.** The encoder (max pooling the preceding layer and without dashed-line box) or decoder (upsampling the preceding layer and concatenating dashed-line box) operation. The boxes indicate the feature maps, with the channel number denoted by equations of $n_{in}$ and $n_{out}$ in the boxes. The number within the arrow indicates the dilation rate.

For each encoder/decoder operation, SDU-Net (one standard convolution and four dilated convolutions) is deeper than U-Net (two standard convolutions) and has a much larger receptive field. Meanwhile, as SDU-Net concatenates the feature maps of all the dilated convolutions to the next encoder/decoder operation, feature maps of the same resolution can be sensed by the network with its multi-scale receptive fields. In addition, for a convolution, the parameter number $n_p = O(n_{ch\_in} \times n_{ch\_out})$, where $n_{ch\_in}$ and $n_{ch\_out}$ indicate the input and output channel numbers. So, if the concatenation output has the same channel number as a single convolution, the four dilated convolutions actually involve much fewer parameters. In general, SDU-Net can ob-tain multi-scale and larger receptive fields using fewer parameters. In our experiment, we set the channel numbers of SDU-Net's encoder/decoder operations as: $n_1 = 64$, $n_2 = 128$, $n_3 = 256$, and $n_4 = 512$.

## 3 Datasets

We adopted four image datasets to test the segmentation performance involving ultrasound images of liver and kidney, breast lesions, thyroid nodules [14], and color digital images of skin lesions [15, 16], as shown in Table 1.

**Table 1.** Datasets. The ratio between training and validation datasets is 8:2.

| Dataset | Train-Val Number | Test Number | Modality |
| --- | --- | --- | --- |
| Thyroid Nodule | 371 | 93 | Ultrasound Image |
| Liver and Kidney | 477 | 116 | Ultrasound Image |
| Breast Lesion | 1237 | 111 | Ultrasound Image |
| Skin Lesion | 2295 | -- | Color Digital Photograph |

The liver-kidney and breast-lesion datasets were retrospectively collected at *Anonymous* after IRB approval and waiver of written consent, and the thyroid-nodule and the skin-lesion datasets are publicly available. Each image of the liver-kidney dataset included both the right liver lobe and the right kidney, so the model was trained and validated for segmentation of both organs at the same time. For the thyroid-nodule dataset, we used the masks preprocessed by a third party [17]. The skin-lesion dataset was obtained from a challenge hosted by the International Skin Imaging Collaboration (ISIC) in 2018. ISIC 2018 did not release the ground truth for the final independent test, so that we did not have independent test data for skin-lesion segmentation, while we had independent test data for the other three tasks.

## 4 Training Procedure

All models were trained from scratch and evaluated using 5-fold cross-validation on the training data, setting the validation split of the data to 0.2. We trained all the networks using a novel loss function, called bi-Dice. The bi-Dice loss function computes the Dice loss with respect to both the object and background, defined as:

$$L_{biDice} = 2 - \frac{2 \sum p_{h,w} \cdot \hat{p}_{h,w} + \epsilon}{\sum p_{h,w} + \sum \hat{p}_{h,w} + \epsilon} - \frac{2 \sum (1 - p_{h,w}) \cdot (1 - \hat{p}_{h,w}) + \epsilon}{\sum (1 - p_{h,w}) + \sum (1 - \hat{p}_{h,w}) + \epsilon}$$

where $(h, w)$ represents the pixel coordinate, $p_{h,w} \in \{0,1\}$ is the mask ground truth, $p_{h,w} = 1$ indicates the pixel belonging to the target, $0 \leq p_{h,w} \leq 1$ is the prediction probability for the pixel belonging to the target, and $\epsilon$ was set to 1 in our experiment.

Adam optimizer was used for training all the networks. The learning rate was set to 0.00005, and all the other parameters were set to the default values in PyTorch. The batch size was set to 4, and the training epochs were set to 500 for all tasks, except for the skin-lesion segmentation. Since the skin-lesion dataset is much larger, we set its training epochs to 85.

## 5    Experiments

We tested the proposed SDU-Net by comparing with vanilla U-Net, AttU-Net, and R2U-NetNet. The comparison included 5-fold cross-validation during the training phase and the test on independent datasets. The Dice score was used as the performance measure, with a higher value indicating better performance. For vanilla U-Net, the input and output channel numbers for each encoder/decoder operation were set the same as SDU-Net. For AttU-Net and R2U-Net, their settings followed the original literature [7, 8], and we adopted the implementation by a third party [18]. The total parameter number of SDU-Net is around 40% of vanilla U-Net's, 17% of AttU-Net's, and 15% of R2U-Net's (Table 2).

**Table 2.** Numbers of parameters for four neural networks.

|        | SDU-Net   | U-Net      | AttU-Net   | R2U-Net    |
|--------|-----------|------------|------------|------------|
| Number | 6,028,833 | 14,787,777 | 34,877,421 | 39,091,265 |

The average Dice scores of the 5-fold cross-validations are presented in Table 3, and the results of the independent test datasets are shown in Table 4. We can see that SDU-Net, vanilla U-Net, and AttU-Net performed well, especially on the liver-kidney dataset and the skin-lesion dataset, while R2U-Net didn't perform well.

**Table 3.** 5-fold cross-validation. The value is shown in mean ± standard deviation of Dice scores. Best performance is highlighted in boldface.

|         | Breast Lesion | Liver/Kidney                  | Thyroid Nodule | Skin Lesion   |
|---------|---------------|-------------------------------|----------------|---------------|
| SDU-Net | **0.843±0.008** | **0.909±0.009/0.810±0.007** | **0.765±0.024** | **0.892±0.003** |
| U-Net   | 0.781±0.020   | 0.896±0.008/0.802±0.012       | 0.722±0.020    | 0.866±0.004   |
| AttU-Net| 0.787±0.014   | 0.900±0.010/0.800±0.010       | 0.705±0.038    | 0.858±0.004   |
| R2U-Net | 0.760±0.033   | 0.661±0.059/0.670±0.032       | 0.496±0.096    | 0.680±0.061   |

**Table 4.** Independent test results. The value is shown in mean ± standard deviation of Dice scores. Best performance is highlighted in boldface.

|         | Breast Lesion | Liver/Kidney                  | Thyroid Nodule |
|---------|---------------|-------------------------------|----------------|
| SDU-Net | **0.859±0.012** | **0.899±0.005/0.797±0.007** | **0.760±0.023** |
| U-Net   | 0.813±0.014   | 0.865±0.020/0.783±0.005       | 0.710±0.017    |
| AttU-Net| 0.816±0.010   | 0.868±0.009/0.787±0.007       | 0.698 ±0.029   |
| R2U-Net | 0.786±0.028   | 0.532±0.088/0.638±0.031       | 0.468±0.086    |

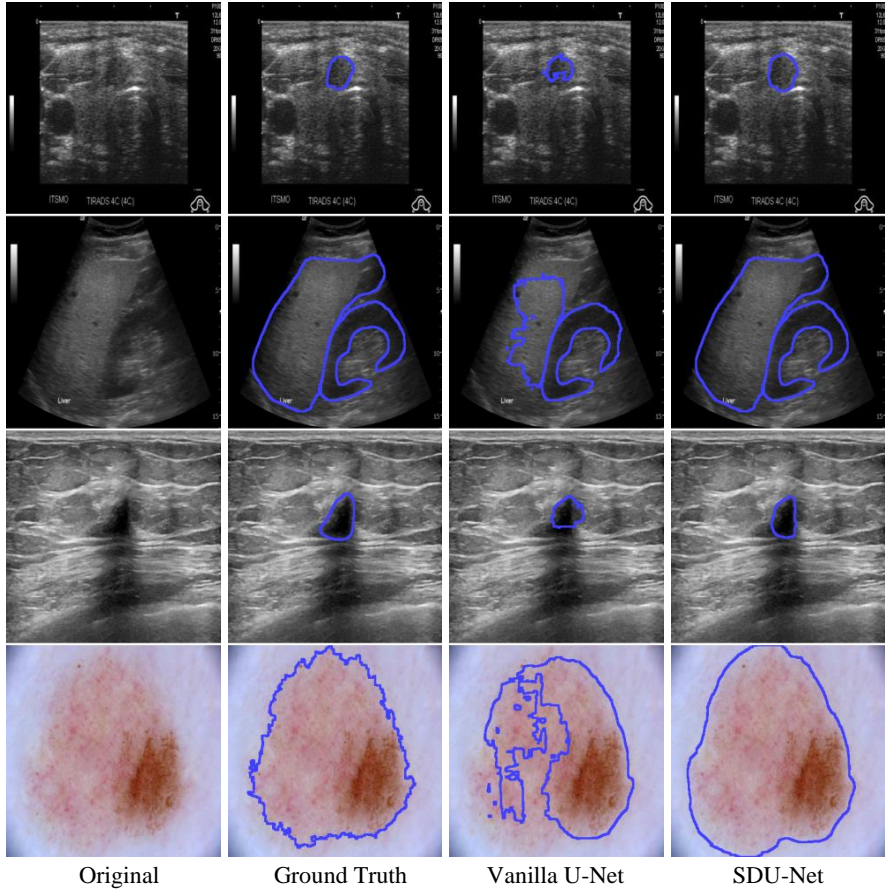

    Original        Ground Truth      Vanilla U-Net      SDU-Net

**Fig. 3.** Segmentation examples. Rows from top to bottom are thyroid nodule, liver & kidney, breast lesion, and skin lesion. Columns from left to right represent original image, ground truth, segmentation result by vanilla U-Net, and segmentation result by SDU-Net. The regions of interest are delineated by blue lines.

Paired t test showed that SDU-Net significantly ($p < 0.05$) outperformed all compared methods on all tasks, except for AttU-Net on thyroid-nodule cross-validation and kidney test. Meanwhile, since SDU-Net used fewer parameters, it was less prone to over-fitting, which was confirmed by the independent test results on the liver-kidney dataset.

Figure 3 demonstrates some examples of segmentation results for each dataset. As depicted in the skin-lesion example, SDU-Net showed higher robustness to local variation and better delineated the whole object. This is in accordance with the fact that SDU-Net has much wider receptive fields. In summary, SDU-Net was more effective and robust, compared to vanilla U-Net, AttU-Net, and R2U-Net, while being a light-weight method for image segmentation.

## 6    Conclusion

This paper introduces a new convolution structure by concatenating the outputs of cascade dilated convolutions. The new segmentation algorithm is computationally more efficient, and has demonstrated higher performance than state-of-the-art algorithms on medical image segmentation datasets, specifically on ultrasound images.